\documentclass[prl,twocolumn,showpacs,floatfix]{revtex4}
\usepackage{epsfig}
\usepackage{amssymb}
\usepackage{amsmath}

\begin{document}

\title[Reentrant Metallic Behavior]{Reentrant Metallic Behavior of
Graphite in the Quantum Limit}

\author{Y. Kopelevich}
\author{J. H. S. Torres}
\author{R. R. da  Silva}
\affiliation{Instituto de F\'{\i}sica "Gleb Wataghin",
Universidade Estadual de Campinas,
Unicamp 13083-970, Campinas, S\~{a}o Paulo, Brasil}
\author{F. Mrowka}
\author{H. Kempa}
\author{P. Esquinazi}
\affiliation{Abteilung  Supraleitung und
Magnetismus, Institut f\"ur Experimentelle Physik II,
Universit\"at Leipzig, Linn{\'e}str. 5, D-04103 Leipzig,
Germany}

\begin{abstract}

Magnetotransport measurements performed on several
well-characterized highly oriented pyrolitic graphite and
single crystalline Kish graphite samples reveal a reentrant
metallic  behavior in the basal-plane resistance at high magnetic
fields, when only the lowest Landau levels are occupied. The
results suggest that the quantum Hall effect and Landau-level-quantization-induced
superconducting correlations are relevant 
to understand the metalliclike state(s) in graphite in the quantum limit.

\end{abstract}

\pacs{71.30.+h, 72.20.My, 73.43.-f, 74.10.+v}
\maketitle

Conduction processes in two-dimensional (2D)
electron (hole) systems, in particular the apparent metal-insulator
transition (MIT) which takes place either varying the carrier
concentration or applying a magnetic field $H$, have attracted a
broad research interest \cite{1}. Recently, a similar MIT driven by a
magnetic field applied perpendicular to basal planes has been reported for
graphite \cite{2,3,4,5}.  The quasi-particles (QP) in graphite behave as 
massless Dirac fermions (DF) with a linear dispersion relation, similar to
the QP near the gap nodes in high-temperature superconductors. Theoretical analysis
\cite{6,7,8} suggests that the MIT in graphite is the
condensed-matter realization of the magnetic catalysis (MC) 
phenomenon \cite{9}
known in relativistic theories of (2 + 1)-dimensional DF. According 
to this theory \cite{6,7,8}, the magnetic
field $H$ opens an insulating gap in the spectrum of DF of
graphene, associated with the electron-hole (e-h) pairing, below a
transition temperature $T_{ce}(H)$ which is an increasing function of
field. However, at higher fields and at temperatures $T < T_{\rm max}(H)$
an insulator-metal transition (IMT) occurs \cite{2} indicating that
additional physical processes may operate approaching the field $H_{QL}$ that pulls
carriers into the lowest Landau level. The occurrence of
superconducting correlations in the quantum limit (QL) \cite{11,12} and
below the temperature $T_{\rm max}(H)$ has been proposed for graphite in
Ref.\cite{2}. On the other hand, authors of Ref.\cite{8} argued
that at high enough carrier concentration,  the basal-plane resistance
$R_b(H,T)$ can decrease decreasing temperature below the e-h pairing
temperature, and identified $T_{\rm max}(H)$ with $T_{ce}(H)$. Other
theoretical works predict the occurrence of the field-induced Luttinger
liquid \cite{13} and the integral quantum Hall effect (IQHE) \cite{14}
in graphite. All these indicate that understanding of the
magnetic-field-induced insulating and metallic states in graphite is of
importance and has an interdisciplinary interest. The aim of this Letter
is to provide a fresh insight on the magnetotransport properties of
graphite in the QL. We show that the IMT is generic to graphite
with a sample-dependent $T_{\rm max}(H)$. Our results of the Hall
resistance $R_h(H,T)$ measurements performed on strongly anisotropic samples 
reveal characteristics related to the QHE.

We have performed measurements of both $R_b(H,T)$ and $R_h(H,T)$
resistances on several well-characterized \cite{2,3,4,5,15,16}
quasi-2D highly oriented pyrolitic graphite (HOPG) and, less anisotropic, 
flakes of single crystalline Kish graphite \cite{17}.
Three HOPG samples with the room temperature and $H = 0$
out-of-plane/basal-plane resistivity ratio $\rho_c/\rho_b = 8.6 \times
10^3$ (HOPG-1) and $\sim 5 \times 10^4$ (HOPG-3 and HOPG-UC), and the Kish
single crystal (K-1) with a ratio of  $\sim 100$ have been studied. HOPG
samples were obtained from the Research Institute ``Graphite", Moscow
(HOPG-1, HOPG-3) and the Union Carbide Co. (HOPG-UC). $\rho_b$ values at
$T = 300$ K $(H = 0)$ are $\sim 3~\mu \Omega$cm (HOPG-UC), $\sim
5~\mu\Omega$cm (HOPG-3, K-1), and $\sim 45~\mu \Omega$cm (HOPG-1).
Low-frequency ($f = 1$ Hz) and dc standard four-probe magnetoresistance
measurements were performed on samples with dimensions $4.9 \times 4.3 \times 
2.5~ \mathrm{mm^{3}}$ (HOPG-1), $4 \times 4 \times 1.2~ \mathrm{mm^{3}}$ (HOPG-3), 
$5 \times 5 \times 1~ \mathrm{mm^{3}}$ (HOPG-UC) and $2.7 \times 2.4 \times 0.15~ \mathrm{mm^{3}}$ 
(K-1) in fields applied parallel to the sample
hexagonal $c$ axis in the temperature interval 70 mK $\le T \le  300$ K
using different 9 T-magnet He cryostats and a dilution refrigerator. The Hall
resistance was measured using the van der Pauw configuration with a cyclic
transposition of current and voltage leads \cite{18,19} at fixed
applied field polarity, as well as magnetic field reversal; no difference
in $R_h(H,T)$ obtained with these two methods was found. For the
measurements, silver past electrodes were placed on the sample surface,
while the resistivity values were obtained in a geometry with a uniform
current distribution through the sample cross section. All resistance
measurements were performed in the Ohmic regime. Complementary
magnetization measurements $M(H,T)$ were carried out with $H || c$ axis
using a SQUID magnetometer.

\begin{figure}
\begin{center}
\epsfig{file=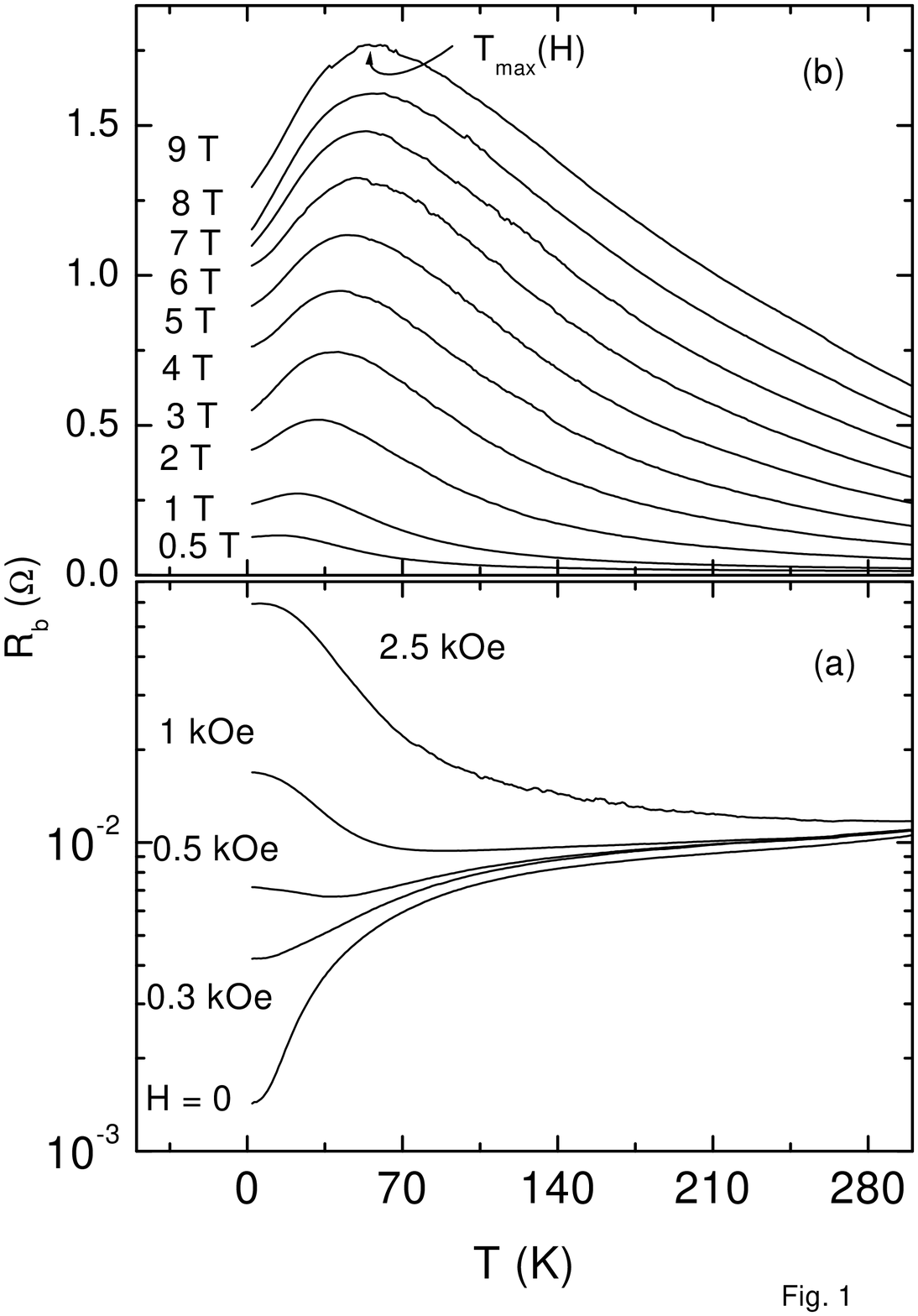,width=\columnwidth}
\end{center}
\vspace{-1cm} \caption[*]{Basal-plane resistance measured in
single-crystalline Kish graphite sample (K-1) in the low- (a) and
high-field (b) regime. Arrow indicates $T_{\rm max}(H)$ below which
reentrant metallic phase appears.}
\label{rb}
\end{figure}

Figure \ref{rb}(a) illustrates the field-induced suppression of the
metallic state measured in the Kish graphite sample (K-1) at low fields,
and Fig. \ref{rb}(b) shows the re-appearance of the metallic state
at $T < T_{\rm max}(H)$ increasing field. We note
that the reentrant metallic state takes place in the Landau level
quantization regime, verified through  measurements of the oscillation in
both $R_b(H)$ and $R_h(H)$ associated with the Shubnikov-de Haas (SdH) effect
\cite{20,21}.

\begin{figure}
\begin{center}
\epsfig{file=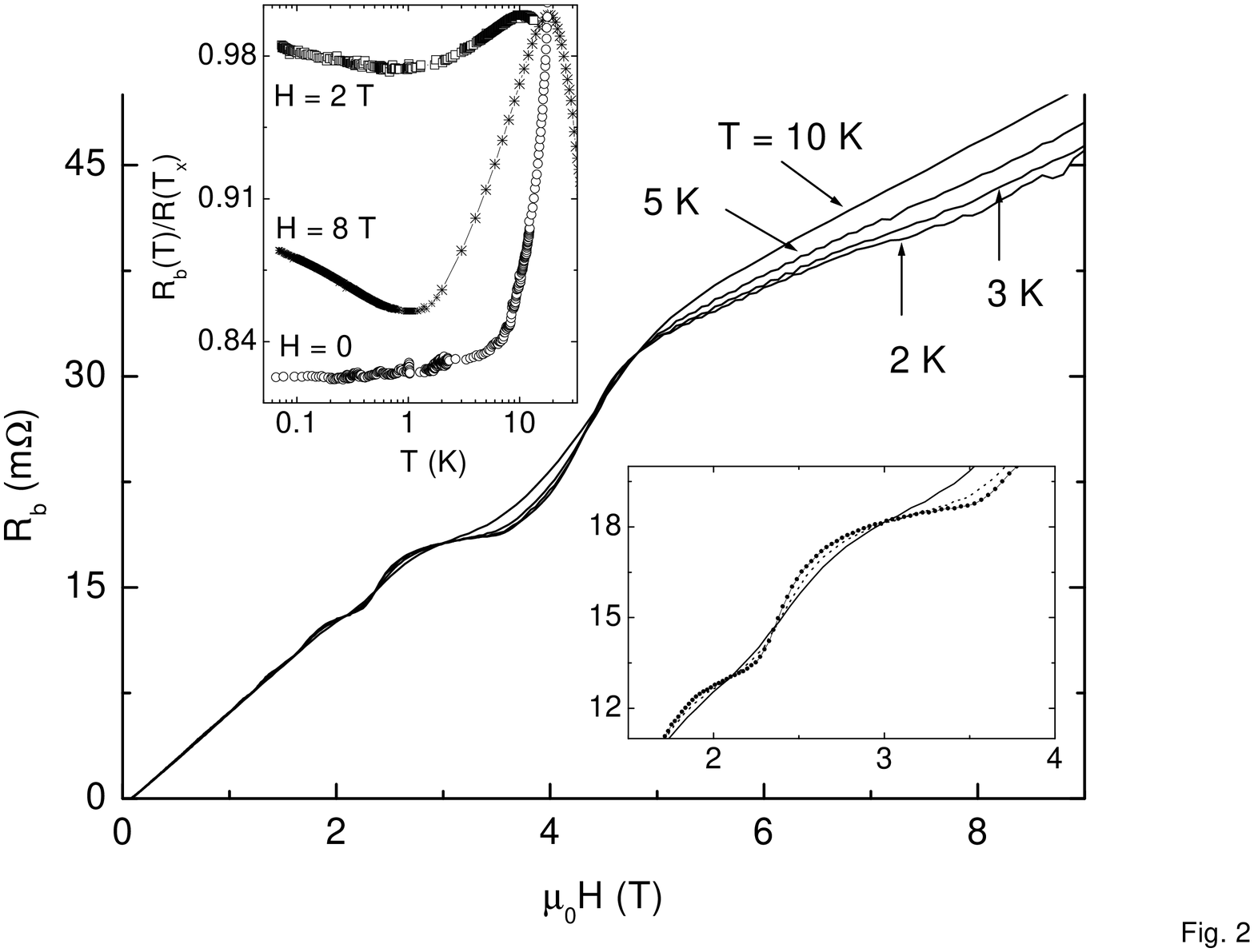,width=\columnwidth}
\end{center}
\caption[*]{Basal-plane resistance measured in HOPG-3 sample at four temperatures,
demonstrating crossings of the $R_b(H)$ isotherms, i.e., the sequence of
the field-driven metal-insulator-metal transitions. The lower inset gives a
detailed view of the crossing in $R_b(H)$ isotherms; $T = 2~$K $(\bullet)$, 5~K (dotted line),
10~K (solid line). The upper inset shows normalized resistance $r=R_{b}(T)/R(T_{x})$ where 
$T_{x}=18~K (H=0, H=8~T)$, and $T_{x}=11~K (H=2~T)$.}
\label{rb2}
\end{figure}

\begin{figure}
\begin{center}
\epsfig{file=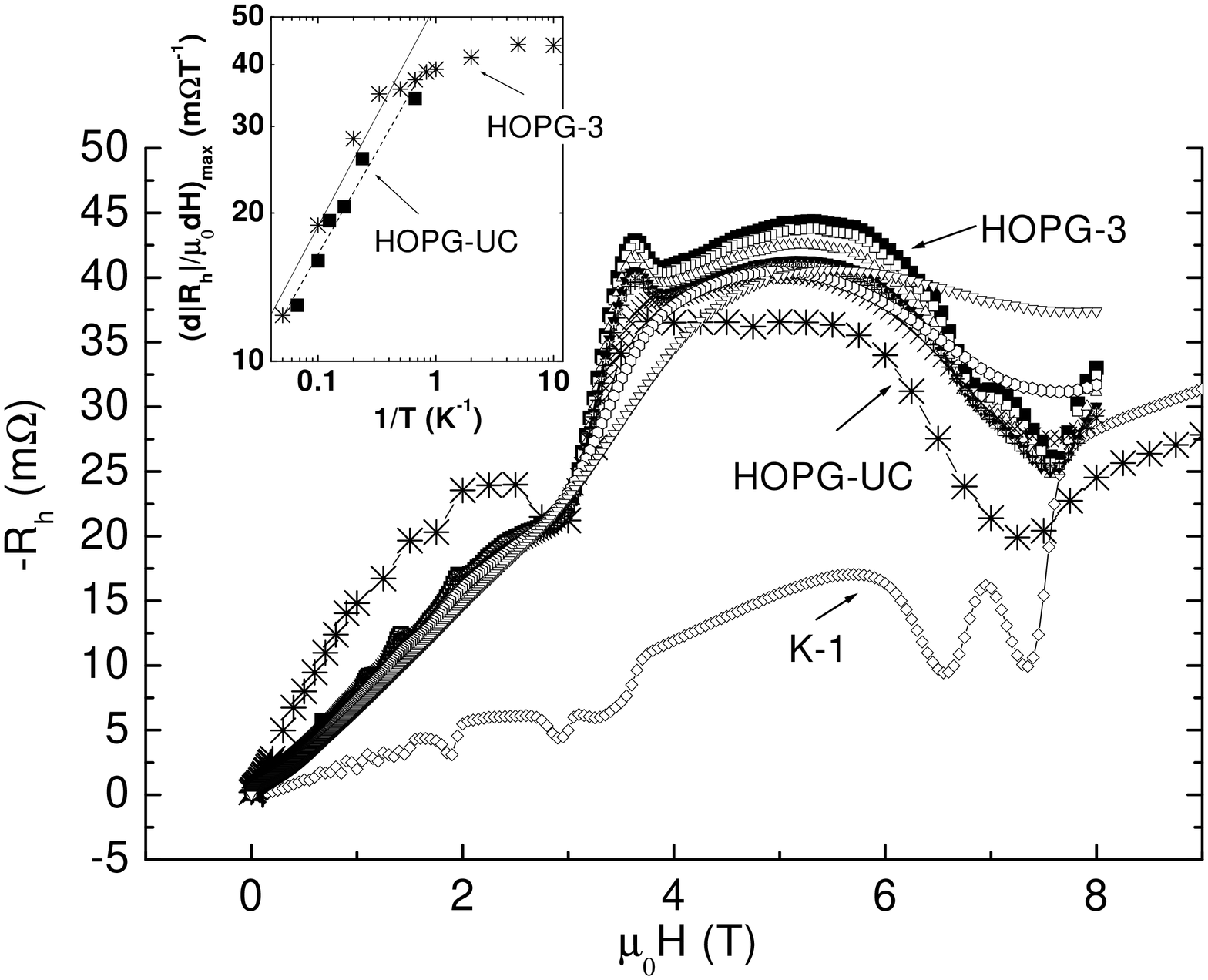,width=\columnwidth}
\end{center}
\caption[*]{Hall resistance $R_h(H,T)$ measured for HOPG-3 sample between
100~mK ($\blacksquare$) to 20~K ($\bigtriangledown)$, for HOPG-UC at T = 4.2~K,
and for K-1 ($R_h/10$) at T = 1.5~K. Inset
shows (d$R_h/$d$H)_{\rm max}$ vs. $1/T$ for the HOPG samples; dashed and
solid lines are linear fits to the function $\sim T^{-\kappa}$ with $\kappa
= 0.42$ and 0.45 for the HOPG samples.} 
\label{rh}
\end{figure}

The noticeable difference in the high-field behavior between HOPG and Kish
graphite samples is a multiple crossing in both $R_b(H,T)$ and $R_h(H,T)$
isotherms measured in HOPG  (see Figs. \ref{rb2} and \ref{rh}) and its absence in
the case of  Kish graphite \cite{21}. The appearance of plateaus-like features in the Hall
resistance $R_h(H)$ measured in HOPG can be seen in Fig. \ref{rh}. The results in
Fig. \ref{rh} suggest the QHE occurrence in HOPG. This new result is not unexpected 
taking into account the quasi-2D nature of HOPG. The QHE has been previously reported 
for various multilayered systems \cite{23,Uji,Zhang} and the occurrence of IQHE in graphene
has been predicted in Ref. \cite{14}.

Following the analysis of transitions between adjacent quantum Hall
plateaus \cite{25}, in the inset of Fig. \ref{rh} we plot the temperature
dependence of the maximum slope (d$|R_h|/$d$H)_{\rm max}$ vs. $T^{-1}$
associated with the largest step in $R_h(H,T)$ measured at $\sim 3.5~$T.
At $T \ge 1.5~$K this slope is $\propto T^{-\kappa}$ with an
exponent $\kappa = 0.42 (0.45)$ for the HOPG-UC
(HOPG-3) sample.  Numerous experiments performed on QHE systems showed
that $\kappa$ varies from sample to sample and can even depend whether it
is determined from Hall or longitudinal resistance measurements
\cite{26,27}. Nevertheless, it is interesting to note that the here obtained
exponent $\kappa$ agrees with that predicted for transitions between both
IQHE and fractional QHE (FQHE) plateaus \cite{28,29}. The observed saturation 
in (d$|R_h|/$d$H)_{\rm max}$ vs. $T^{-1}$ at $T < 1.5~$K (see inset in
Fig. \ref{rh}) is similar to that found in QHE systems but its origin is
still unclear \cite{30,31,32}. The analogous behavior has also been reported for
other quasi-2D bulk QHE systems as, e. g., $(TMTSF)_{2}AsF_{6}$ \cite{Uji}. We 
stress that $R_{h}(H,T)$ is the Hall resistance
measured for the bulk sample which translates, e. g., to $\rho_{h}=3.5~m\Omega cm$ at
the main plateau for HOPG-UC sample. This gives $R_{h}/\square = \rho_{h}/d \sim 10~k\Omega$
 ($d=3.35$ \AA~is the interlayer distance), i. e., only a factor $\sim 2.5$ less than
the Hall resistance quanta $h/e^{2}$. The upper inset in Fig. \ref{rb2}
shows $R_{b}(T)$ measured at H = 0, 2, and 8 T down to 70 mK, the lowest available 
temperature. As can be seen, at high fields and $T < 1~K$ the resistance drop is followed
by a logarithmic increase which can be accounted for by a possible formation of the
Wigner crystal or charge density wave of Cooper pairs (see below) in quasi-2D systems \cite{38,39}
in the presence of quenched disorder.

We note further that if the QHE-like behavior of HOPG samples is related to their 
quasi-2D nature, the lack of any signature for the QHE in Kish graphite
provides an additional evidence for its 3D character. On the other hand,
the reentrant metallic state takes place for all filling factors
or magnetic fields $H > H_{QL}\sim 4$ T for HOPG samples, and $\mu_0 H > 0.2$ T for
the K-1 sample, indicating that the QHE alone cannot account for this effect.

\begin{figure}
\begin{center}
\epsfig{file=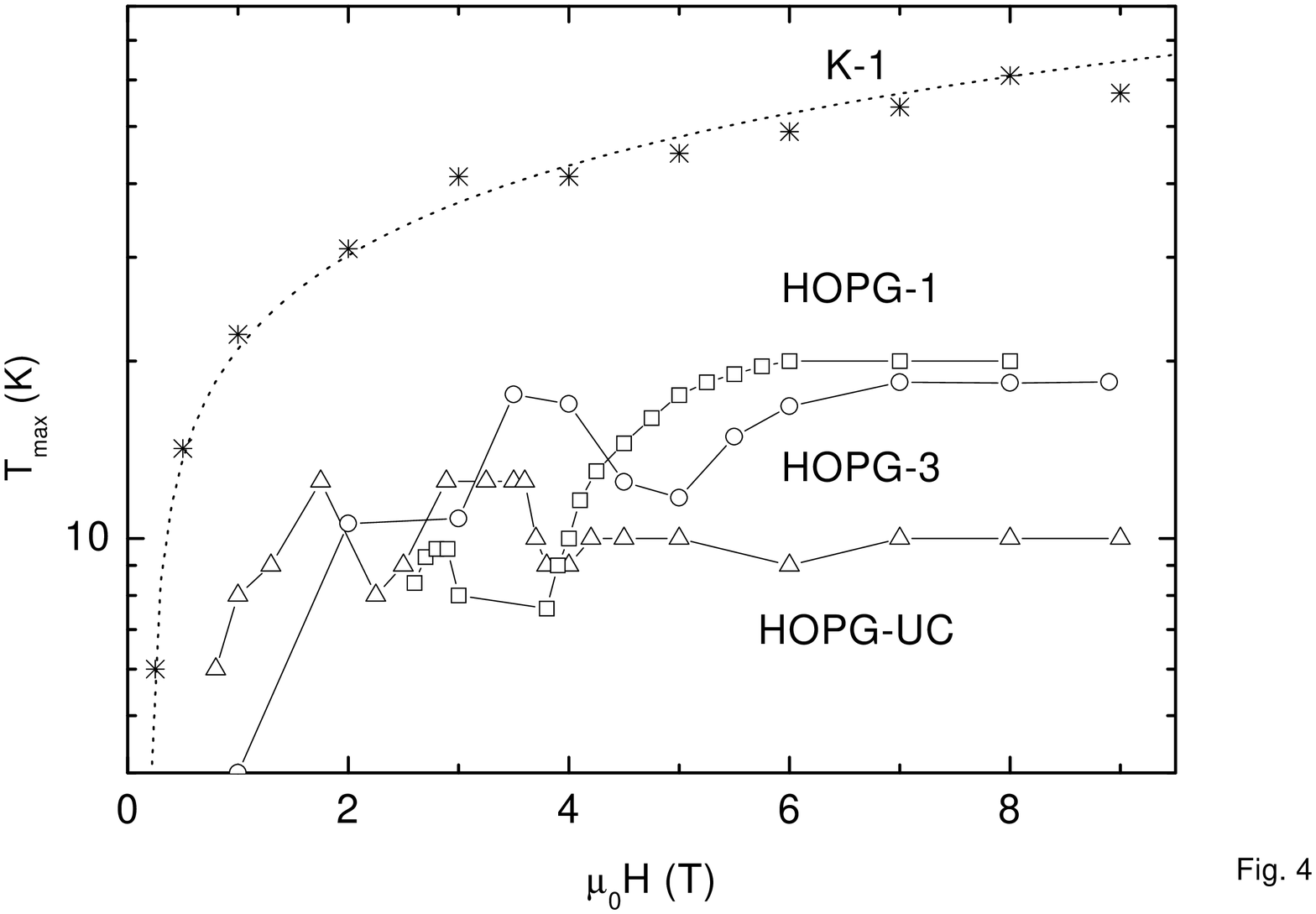,width=\columnwidth}
\end{center}
\caption[*]{$T_{\rm max}$ vs. $H$ for all studied samples. Dotted line
corresponds to Eq.(1) with $C = 21.5$~K$/$T$^{0.5}$ and $\mu_0 H_c =
0.16~$T.} 
\label{tmax}
\end{figure}

Figure \ref{tmax} presents $T_{\rm max}(H)$ obtained for all measured
samples. It demonstrates that $T_{\rm max}(H)$ for the K-1 sample can be 
surprisingly well approximated by the equation
\begin{equation}
T_{\rm max}(H) = C [1 -  (H_c/H)^2]H^{1/2}\,, \label{eqtmax}
\end{equation}
with $H_c$ and $C$ being fitting parameters. The Eq. (1) is similar to
the expression (64) of Ref. \cite{8}, $T_{ce} \sim (1 -
\nu_b^2)H^{1/2}$, obtained within the MC theory, where $\nu_b = 2\pi c
n_{2D}/N_f|eH| \equiv H_c/H$ is the filling factor, $N_f$ is the number of
fermion species ($N_f = 2$ for graphite), and $n_{2D}$ is the 2D carrier
density. We note, however, the order of magnitude difference between the
predicted value for $\mu_0 H_c \approx 2.5~$T (using $n_{2D} = n_{3D}d \sim
10^{11}~$cm$^{-2}$ taking $n_{3D} \sim  3 \times 10^{18}$~cm$^{-3}$) \cite{8} and the
fitting value $\mu_0 H_c = 0.16~$T; see Fig. \ref{tmax}.

Alternatively, the reentrant metallic state in both HOPG and Kish graphite
samples can be caused by a common mechanism associated with the Cooper-pair
formation. The appearance or reappearance of superconducting correlations
in the regime of Landau level quantization has been predicted by several
theoretical groups (for review articles see Refs. \cite{11,12}). According to the theory,
superconducting correlations in quantizing field result from the increase of 
the 1D density of states $N_1(0)$ at the Fermi level. In the quantum limit
$(H > H_{QL})$ the superconducting critical temperature $T_{SC}(H)$ for a 3D system 
is given by the equation \cite{11}
\begin{equation}
T_{SC}(H) = 1.14 \Omega \exp({-2\pi l^2/N_1(0)V})\,, \label{tc}
\end{equation}
where $2\pi l^2/N_1(0) \sim 1/H^2, l = (\hbar c/eH)^{1/2}, V$ is the
BCS attractive interaction, and $\Omega$ is the energy cutoff on $V$ (in 2D case, 
$T_{SC}$ increases linearly with field \cite{12,34,35}). The increase
of $T_{\rm max}$ with field (see Fig. \ref{tmax}) is in a qualitative
agreement with Eq. (\ref{tc}) and the 2D predictions \cite{12,34,35}. Above a certain
field $H > H_{QL}$ a reentrant decrease of $T_{SC}$  is also expected \cite{11}, being 
consistent with the saturation in $T_{\rm max}(H)$; see Fig. \ref{tmax}. The occurrence of
either spin-singlet or spin-triplet \cite{36} superconductivity in graphite may be possible
in the QL. Theory predicts an oscillatory behavior of $T_{SC}(H)$
at $H < H_{QL}$, i.e., with an increasing number of occupied Landau levels; indeed, a nonmonotonic
$T_{\rm max}(H)$ is observed for all HOPG samples at $\mu_0 H < 4~$T (see Fig.~\ref{tmax}). The
absence of pronounced $T_{\rm max}$ vs. $H$ oscillations in Kish graphite can naturally be understood taking 
into account its lower anisotropy. In (quasi-)2D case the density of states $N(0)$ is a 
set of delta functions (broadened however by quenched and thermal disorder) corresponding 
to different Landau levels, and hence $T_{\rm max}$ should oscillate stronger with field 
in HOPG, as observed. A $T_{\rm max}(9$T$) = 62$~K obtained for Kish graphite is much higher 
than $T_{\rm max}(9$T)$~= 11$~K measured for strongly anisotropic HOPG-UC sample. This 
fact can be understood taking into account quantum and/or thermal fluctuations \cite{11,12}, 
which are stronger in quasi-2D HOPG and hence can effectively reduce $T_{\rm max}$ (i.e., $T_{SC}$). 
It is expected that below $T_{SC}(H)$ and for 3D samples, the resistance along the applied 
field vanishes and the resistance perpendicular to the field direction shows a drop \cite{11,12}. 
However, in graphite both the $c$-axis and basal-plane resistances remain finite due to the 
layer crystal structure, implying the occurrence of superconducting correlations without 
macroscopic phase coherence.

\begin{figure}
\vspace{+1cm}
\begin{center}
\epsfig{file=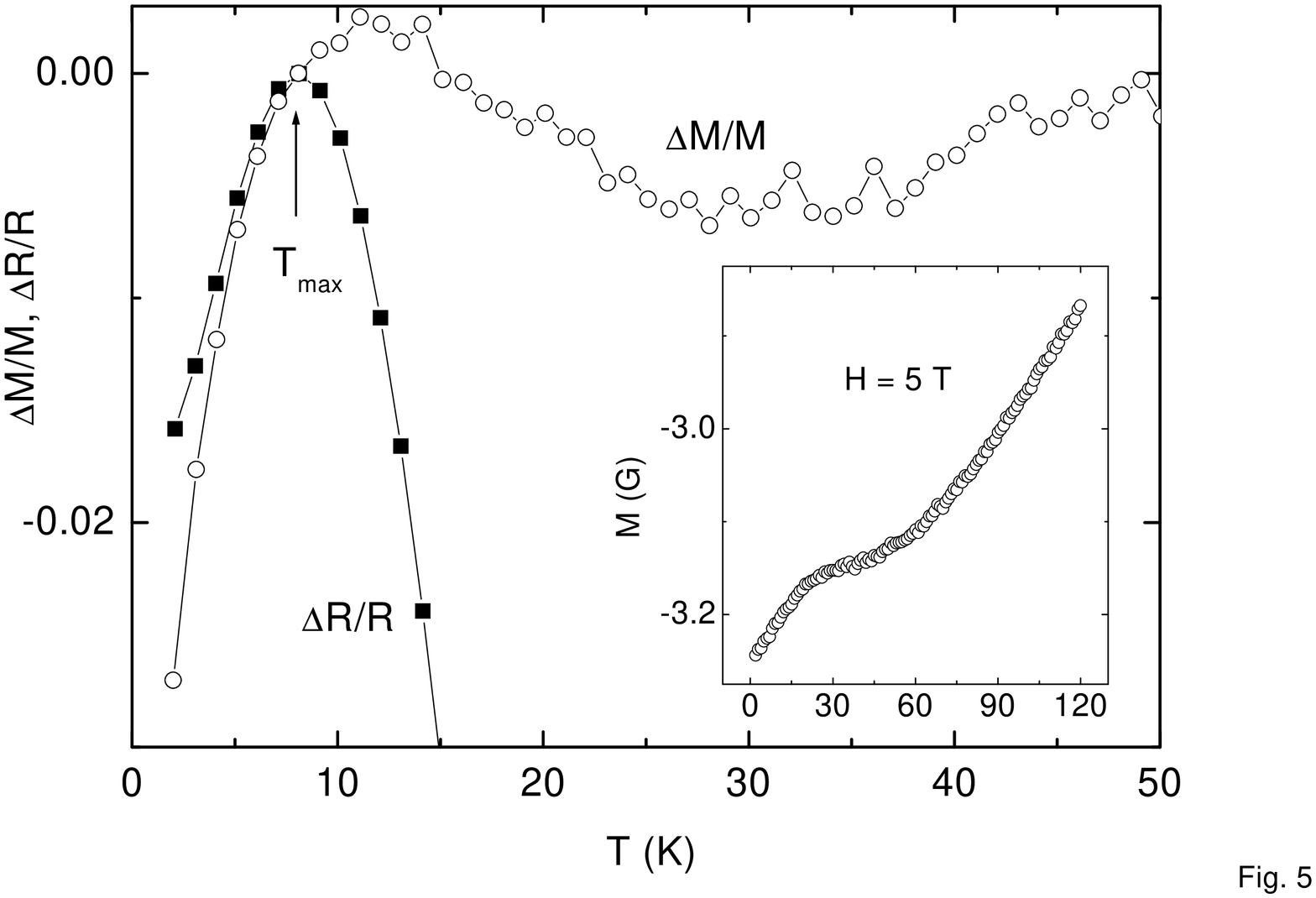,width=\columnwidth}
\end{center}
\caption[*]{$\Delta M/M = [M(T) - M(T_{\rm max})]/ M(T_{\rm max})$
and $\Delta R/R = [R_b(T) - R_b(T_{\rm max})]/ R_b(T_{\rm max})$
measured for the HOPG-UC sample at $\mu_0 H = 1~$T. The inset presents
the magnetization $M(T)$ measured at $\mu_0 H = 5~$T.}
\label{mag}
\end{figure}

In Fig. \ref{mag} we compare $R_b(T)$ and the magnetization $M(T)$
measured  for the HOPG-UC sample at $\mu_0 H = 1~$T and 5~T (inset), viz.,
in the QL, illustrating that the reentrant metallic state(s) is(are)
accompanied by the enhanced diamagnetic response, supporting both the
superconductivity- and QHE-based scenarios for the field-induced metallic
state(s). 

In summary, the results of the present work provide an unambiguous evidence for
the reentrant metallic state(s) in graphite induced by a magnetic field in
the quantum limit. The observed sequence of the field-driven insulator-metal-insulator
transitions as well as a signature for the QHE in HOPG samples rule out any trivial 
explanation for the phenomenon. On the 
other hand, the overall results can consistently be understood assuming the 
occurrence of superconducting correlations in the regime of Landau level quantization,
providing thus a possible solution of the longstanding problem of the metallic
resistance behavior $(dR_{b}/dT>0)$ in graphite in the QL even below 1K \cite{37}.

\begin{acknowledgments}
This work was supported by the following institutions and grants: FAPESP,
CNPq, CAPES, DAAD and DFG  under Grant ES 86/6-3.
We gratefully acknowledge helpful discussions with D. V. Khveshchenko, I.
A. Shovkovy, V. A. Miransky, and V. P. Gusynin.

\end{acknowledgments}

\end{document}